\documentclass[onecolumn,preprintnumbers, superscriptaddress, reprint, longbibliography, prl]{revtex4-1}
\usepackage{amsmath}
\usepackage{stmaryrd}
\usepackage{txfonts}
\usepackage{amssymb}
\usepackage{mathrsfs}
\usepackage{graphicx}
\usepackage{dcolumn}
\usepackage{bm}
\usepackage{braket}
\usepackage{epsfig}
\usepackage{color}
\usepackage{ulem}
\usepackage{bibunits}

\usepackage[colorlinks=true, linkcolor=blue, citecolor=blue, urlcolor=blue]{hyperref}
\begin{document}

\title{Topological Sliding Moir{\' e} Heterostructure}
\author{Ying Su}
\affiliation{Theoretical Division, T-4 and CNLS, Los Alamos National Laboratory, Los Alamos, New Mexico 87545, USA}
\author{Shi-Zeng Lin}
\affiliation{Theoretical Division, T-4 and CNLS, Los Alamos National Laboratory, Los Alamos, New Mexico 87545, USA}

\begin{abstract}
We investigate the effect of sliding motion of layers in Moir{\'e} heterostructures on the electronic state. 
We show that the sliding Moir{\'e} heterostructure can {generate} nontrivial topology characterized by the first and second Chern number in the high dimensional manifold spanned by the physical dimensions and {synthetic dimensions associated with the sliding displacement}.
The nontrivial topology implies a {topological} charge pumping caused by the sliding motion. We demonstrate the nontrivial topology and charge pumping explicitly in a one dimensional bi-chain model and the {small}-angle twisted bilayer graphene. 
{Contrary to the conventional belief that the interlayer sliding in {incommensurate} Moir{\'e} Heterostructures does not affect {the} {electronic structure}, our results reveal  that the sliding motion can generate nontrivial topology dynamically and hence cannot be neglected}  {in the dynamical process.}

\end{abstract}

\date{\today}
\maketitle
The Moir{\'e} heterostructure (MH) obtained by stacking two dimensional (2D) materials have attracted considerable interest recently. The Moir{\'e} pattern emerges when the underlying 2D materials have different lattice constants or 
twisted crystal orientations, and results in a periodic spatial modulation. \cite{geim_van_2013,hunt_massive_2013,dean_hofstadters_2013,ponomarenko_cloning_2013} The Moir{\'e} potential with modulation period much larger than the atomic lattice constant has profound effect on the single electron band structure. For example, the kinetic energy of electrons can be quenched {significantly} by interlayer twisting which effectively enhances the correlation between electrons \cite{PhysRevLett.99.256802,bistritzer_moire_2011}.  As a consequence, novel quantum states enabled by strong correlation can be stabilized in {MHs}. Recently superconductivity, {correlated insulator,} and quantum anomalous Hall state have been observed in magic-angle twisted graphene multilayers \cite{cao_correlated_2018,cao_unconventional_2018,Yankowitz1059,lu2019superconductors,chen2019evidence,
chen2019Signatures,shen2019observation,liu2019spin,cao2019electric,Sharpe605, serlin2019intrinsic}.

{Interlayer twisting and sliding are two internal degrees of freedom in MHs. {Especially in the incommensurate MHs, the translational motion of one layer with respect to the other layers does not cost energy and is a Goldstone mode.} 
The {rigid} sliding displacement shifts the Moir{\'e} pattern and {is expected not to} change the electronic band structure {for incommensurate MHs} \cite{bistritzer_moire_2011}}. Hence it is usually neglected in studying the electronic properties of MHs.

\begin{figure}[t]
  \begin{center}
  \includegraphics[width=8.5 cm]{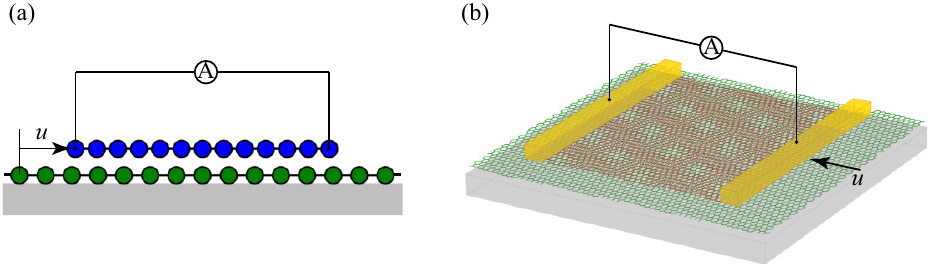}
  \end{center}
\caption{Schematic setup for measuring the topological pumping in sliding bi-chain (a) and sliding twisted bilayer graphene (b).
} 
  \label{fig1}
\end{figure}

In this paper we point out that the sliding of the layer in MHs has dynamical effects on the electronic structure. 
{We show that the sliding displacement configures synthetic dimensions with which the Hamiltonian manifold is transcendent to higher dimensions.}
For example, the effective dimension of sliding 1D (2D)  MHs
becomes two (four)  after including 
{the synthetic dimensions associated with the sliding displacement in 1D (2D) real space. In the high dimensional manifold, the nontrivial topology can be generated dynamically even though the electronic state is topologically trivial in  real dimensions.} 
The topology of the electronic state is characterized by the first (second) Chern number defined in the 
2D (4D) manifold. {Because the Chern numbers associated with the synthetic dimension are  generated dynamically, we call them dynamical Chern number (DCN) to be distinct from the conventional Chern number defined in real dimensions. Due to} the nontrivial topology, there is topological pumping of electrons during the sliding motion of layers which can be {measured} in the setup sketched  in Fig. \ref{fig1}. 


{We consider two MHs, the 1D bi-chain model and the {small}-angle twisted bilayer graphene (TBG), to illustrate the nontrivial topology generated by interlayer sliding. In the bi-chain model, electrons hopping on two coupled 
1D chains whose lattice constants are $a_l$ (where $l=1,2$ for the top and bottom chain) are described by} 
\begin{equation}
\mathcal{H}=-\sum_{\langle ij\rangle,l} t c_{i,l}^\dagger c_{j,l} + \sum_{i,j} \left[ t_{ij}(u_1-u_2)c_{i,1}^\dagger c_{j,2} +\text{H.c.} \right],
\end{equation}
where $u_1$ ($u_2$) is the sliding distance of the top (bottom) chain.
Here the first term describes the nearest neighboring intrachain hopping and the last term is the interchain tunneling. The tunneling strength is 
assumed to be $t_{ij}(u_1-u_2)=t'e^{-|r_{i,1}+u_1-r_{j,2}-u_2|/\lambda}$ with $r_{i,l}$ being the position of the $i$-th site of the $l$-th chain without sliding. {The sliding motion is much slower than the typical electronic dynamics, therefore can be regarded as an adiabatic process.} 
Sliding one chain by its lattice constant does not change the system. Therefore $\mathcal{H}$ is a periodic function of $u_l$ with period $a_l$. This can be seen clearly by representing the interchain tunneling in the basis of Bloch states $c_{k,l}={\frac{1}{\sqrt{N_l}}}\sum_je^{-i k {(r_{j,l}+u_l)}} c_{j,l}$ as $\sum_{i, j}t_{ij}(u_1-u_2)c_{i,1}^\dagger c_{j,2}=\frac{1}{\sqrt{a_1a_2}}\sum_{G_1, G_2, q} t_q' e^{i(G_1u_1+G_2u_2)} c_{q-G_1, 1}^\dagger c_{q+G_2, 2}$, where $t_q'$ is the Fourier transform of $t_{ij}(u_1-u_2)$  and $G_l$ ($N_l$) is the reciprocal wavevector (number of sites) of the $l$-th chain. Because $\mathcal{H}$ has different periods when sliding different chains,  the resulting topology is also different for shifting different chains. To facilitate the discussion, we slide the top chain only with $u_1=u$ and $u_2=0$ in the following (see Fig. \ref{fig1}(a)). 

\begin{figure}[t]
  \begin{center}
  \includegraphics[width=8.5 cm]{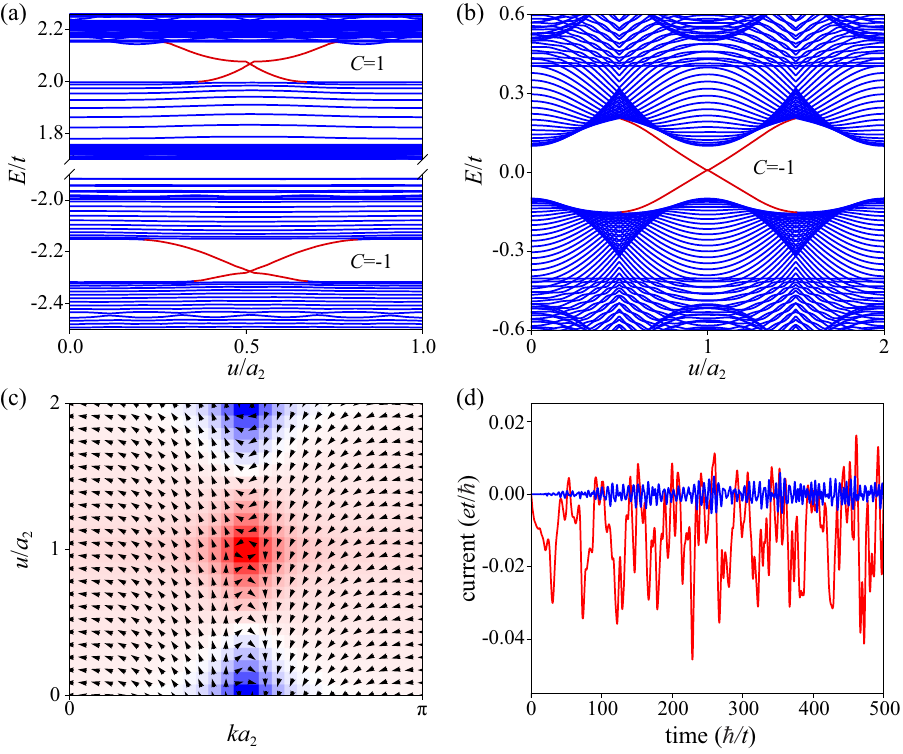}
  \end{center}
\caption{Energy spectrum of the incommensurate bi-chain model (a) and commensurate bi-chain model (b). (c) The spin texture of $\mathbf{S}(k,u)$ for the effective Hamiltonian of bi-chain model Eq. \eqref{eq3a}. The arrows represent the direction of in-plane components and the color (with red for positive and blue for negative) stands for the $z$ component of  $\mathbf{S}(k,u)$. (d) The currents in the top chain (red line) and bottom chain (blue line) for the sliding bi-chain model. } 
  \label{fig2}
\end{figure}

To be concrete, we consider the case where the bottom chain is infinitely long while the top chain is finite and is much longer than the lattice constant. We treat the bottom chain as substrate and apply periodic boundary condition to it in the calculation. Now the bi-chain model has the symmetry $\mathcal{H}(u)=\mathcal{H}(u+a_2)$.  To explore how the slide of the top chain changes the electronic states, we study the electronic spectrum as a function of $u$. Here we fix $a_1/a_2=\sqrt{5}-1$ such that the two chains are incommensurate in general and form a Moir{\'e} pattern with the period of $a_M=a_1a_2/|a_1-a_2|$. For $t'=0.8t$ and $\lambda=0.5a_2$, the electronic spectrum is shown in Fig. \ref{fig2}(a). Apparently, there are bulk energy gaps and edge states traversing the gaps that exhibits the signature of topological insulator. 
{The edge states are localized at the two ends of the top chain and its appearance depends on $u$. To characterize the nontrivial topology, we employ the twisted boundary condition to the top chain $c_{i+N_1,1}=e^{i\xi}c_{i,1}$ where $\xi$ varies from 0 to $2\pi$ \cite{PhysRevLett.97.036808}. Here $(\xi, u)$ spans a compact 2D manifold and the DCN is defined as    
$C^{\mu\nu} =\frac{1}{2\pi} \sum_{E_n\le E_F}  \int d \xi d u \Omega_n^{\mu\nu}$
where $ \Omega_n^{\mu\nu}= i\braket{\partial_\nu \phi_{n}| \partial_\mu\phi_n} - i \braket{\partial_\mu\phi_n| \partial_\nu \phi_n}$ is the Berry curvature, $\ket{\phi_n}$ is the eigenstate of the $n$-th band {with eigen energy} $E_n$, and $\mu,\nu=\xi,u$ in this case. For the Fermi energy in the top (bottom) bulk energy gap in Fig. \ref{fig2}(a), the Chern number is $C^{\xi u}=1$ ($-1$) that is consistent with the {number of pairs of edge states.}}

To gain further insights into the topological nature of the sliding bi-chain model, we consider a simplified version by setting $a_1/a_2=1$ and introducing a staggered on-site potential $\Delta_{i,2}=(-1)^i\Delta$ to the bottom chain. The staggered potential doubles the period of the system $\mathcal{H}(u)=\mathcal{H}(u+2a_2)$ and each unit cell contains four different sites. {The Hamiltonian is
\begin{equation}
\mathcal{H} =
\begin{pmatrix}
h_{1}(k) & T(k,u) \\
T^\dagger(k,u) &  h_{2}(k)   
\end{pmatrix},  
\end{equation}
where $h_1(k)=-2t\cos(ka_2)\sigma_x$ and $h_2(k)=-2t\cos(ka_2)\sigma_x  + \Delta\sigma_z$ are for the top and bottom chains, respectively. The interchain coupling is 
$T_{\alpha\beta} = \frac{1}{2a_2}\sum_G e^{iG(\delta_\alpha+u-\delta_\beta)} t_{k+G}'$
where $\alpha$ and $\beta$ denote different sublattices with the internal position $\delta_\alpha$ and $\delta_\beta$ in a unit cell.} The electronic spectrum in this case for $\Delta=0.4t$ and $t'=0.5t$ is shown in Fig. \ref{fig2}(b) in which the topological edge states emerge around $E=0$. The staggered potential gaps out the electronic states near $E=0$ of the bottom chain. 
This allows us to treat the second chain as a perturbation to the first chain for a weak interchain coupling near $E=0$. The effective Hamiltonian for the top chain becomes
\begin{equation}\label{eq3a}
\mathcal{H}_1(k,u) = h_1(k) - T(k,u) h_2^{-1}(k)T^\dagger(k,u)=E_0+\mathbf{S}\cdot\bm{\sigma},
\end{equation}
{which is a $2\times2$ matrix expressed in terms of Pauli matrices $\bm{\sigma}$ acting on sublattice space.}
The texture of the $\mathbf{S}$  vector is displayed in Fig. \ref{fig2} (c) with nontrivial topology characterized by the skyrmion number $C=\frac{1}{4\pi}\int dk du \mathbf{S}\cdot(\partial_k \mathbf{S}\times \partial_u \mathbf{S})=-1$ \cite{qi2006}, which is equivalent to the DCN and consistent with the one calculated using the twisted boundary condition.

The 1D bi-chain model clearly demonstrates the sliding motion generates nontrivial topology. 
We next turn to the TBG system, which is more relevant to experiments.

\begin{figure}
  \begin{center}
  \includegraphics[width=8.5 cm]{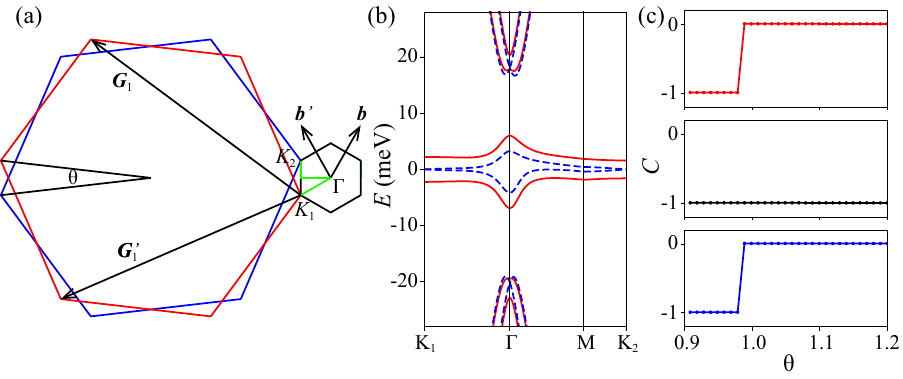}
  \end{center}
\caption{(a) Brillouin zones for the top layer graphene (red), bottom layer graphene (blue), and Moir{\'e} superlattice (black). (b) Band structure of the magic-angle TBG with sublattice potential (red solid line) and without sublattice potential (blue dashed line) along the high symmetry path in the MBZ. (c) Chern numbers as a function of the twist angle $\theta$. The bottom to top panels correspond to the Fermi energy in the three band gaps from bottom to top in (b). } 
  \label{fig3}
\end{figure}

{In TBG, the single-particle band structure at a small twisted angle can be described satisfactorily by a continuum model \cite{bistritzer_moire_2011,PhysRevB.86.155449}. In this model, only the electronic state near the Dirac cones of two graphene layers and the small momentum transfer between layers due to the smooth Moir{\'e} potential are considered. Generally, the MH is not exactly periodic. However, the 
atomic-scale incommensurability is neglected in the continuum model because 
 of the lattice relaxation and fast decay of tunneling amplitude with momentum transfer. %
 Thus the interlayer coupling potential can be expanded to the lowest order respecting the Moir{\'e} pattern symmetry. Here we employ the continuum model to describe the TBG as \cite{bistritzer_moire_2011}}
\begin{equation}\label{HTBG}
\mathcal{H}_\tau =
\begin{pmatrix}
h_{1,\tau}(\bm{k}) & T_\tau(\bm{r},\bm{u}) \\
T_\tau^\dagger(\bm{r},\bm{u}) &  h_{2,\tau}(\bm{k})   
\end{pmatrix},  
\end{equation}
where $\tau=\pm$ is the valley index and $\mathcal{H}_\pm$ are related by time-reversal symmetry (TRS). $h_{l,\tau}(\bm{k})$ is the low-energy effective Dirac Hamiltonian of the $l$-th layer graphene and at the $\tau K$ valley.  Under the twist that the bottom (top) layer is rotated by $\theta/2$ ($-\theta/2$), the Dirac Hamiltonian becomes 
\begin{equation}
h_{l,\tau}(\bm{k}) =\mathcal{R}(\pm\tau\theta/2) \hbar v_F(\bm{k}-\tau\bm{K}_l)\cdot \bm{\sigma}_{\tau} \mathcal{R}(\pm\tau\theta/2)^{-1},
\end{equation}
where $\mathcal{R}(\theta)=e^{-i\theta\sigma_z/2}$ is the rotation operator  and $\bm{\sigma}_\tau=(\tau\sigma_x, \sigma_y)$ is the Pauli matrix acting on sublattice space. Here $\pm\theta/2$ are respectively for $l=1,2$ and $\bm{K}_l$ are the corners of Moir{\'e} Brouillon zone (MBZ) as shown in Fig. \ref{fig3}(a). The interlayer tunneling is described by 
\begin{equation}
\begin{split}\label{eq12}
&T_\tau(\bm{r},\bm{u})=T_\tau^{(0)} + e^{-i\tau(\bm{b}\cdot\bm{r}+\bm{G}_1\cdot\bm{u})} T_\tau^{(1)} + e^{-i\tau(\bm{b}'\cdot\bm{r}+\bm{G}_1'\cdot\bm{u})} T_\tau^{(2)}, \\
&T_\tau^{(n)}=w_{AA}\sigma_0+w_{AB}\cos\left(\frac{2\pi n}{3}\right)\sigma_x+\tau w_{AB}\sin\left(\frac{2\pi n}{3}\right)\sigma_y,
\end{split}
\end{equation}
where $\bm{b}$ ($\bm{b}'$) and $\bm{G}_1$ ($\bm{G}_1'$)  are the primitive reciprocal lattice vectors of the MBZ and of the top layer graphene BZ, respective (see Fig. \ref{fig3}(a)). Due to the lattice relaxation, the interlayer distance at AA stacking regions is larger than that at AB stacking regions. Thus there are two different interlayer tunneling strengths $w_{AA}=79.7$ meV and $w_{AB}=97.5$ meV \cite{Fuprx}. The Fermi velocity is $\hbar v_F/a=2.1354$ eV where $a=0.246$ nm is the lattice constant of single layer graphene. $\bm{u}$ is the sliding displacement of the top layer with respect to the bottom layer, which causes shift of the Moir{\'e} pattern. Therefore shifting $\bm{u}$ does not change the electronic band structure. The phase factor associated with $\bm{u}$ in Eq. \eqref{eq12} can be gauged away by changing the basis for the bottom layer $\ket{\bm{k}}\rightarrow \exp(i \varphi_k)\ket{\bm{k}}$ with $\varphi_k=n \bm{G}_1 \cdot\bm{u}+m  \bm{G}_1'\cdot\bm{u}$ and $m, \ n$ are integers \cite{bistritzer_moire_2011}. {However, for sliding TBG with fixed gauge for the basis, the $\bm{u}$ dependent phase should be kept. }

The band structure of TBG with the magic angle $\theta = 1.05^\circ$ and {of the $K$ valley} is shown in Fig. \ref{fig3}(b) where there are two flat bands (without considering the  degenerate spin degree of freedom) near the Fermi energy {and touching at the MBZ corners due to the $\mathcal{C}_2\mathcal{T}$ and $\mathcal{C}_3$ symmetries \cite{PhysRevX.8.031089}. Here $\mathcal{C}_n$ and $\mathcal{T}$ are the $n$-fold rotation and time-reversal operators.} The flat bands are well separated from remote bands due to lattice relaxation.
They can be gapped by breaking the $\mathcal{C}_2$ symmetry, i.e. by placing TBG on hexagonal boron nitride (hBN) substrate. The hBN substrate induces a sublattice potential $\Delta_{AB}$ on the bottom layer graphene that results in gapped flat bands as shown in Fig. \ref{fig3}(b) for $\Delta_{AB}=30$ meV. In this case, the flat bands have nonzero Chern numbers $C=\pm 1$ for the top and bottom flat bands, respectively \cite{PhysRevB.99.075127}. The flat bands of the $-K$ valley have opposite Chern numbers due to the TRS.  When there is a valley polarization due to interaction, the TRS is spontaneously broken, which results in quantized anomalous Hall effect, and has been observed experimentally \cite{Sharpe605, serlin2019intrinsic}. {The nontrivial topology of flat bands can be tuned by changing the twist angle $\theta$. Here we focus on a small range of $\theta$ around the magic angle where the continuum model is valid,  the Chern numbers for the Fermi energy in the three band gaps (from bottom to top) in Fig. \ref{fig3}(b) are shown in Fig. \ref{fig3}(c) (bottom to top panels). Apparently, there is a jump of Chern numbers from $-1$ to 0 for the Fermi energy in the top and bottom gaps at a critical angle $\theta_c=0.98^\circ$. However,  for the Fermi energy in the middle gap, the Chern number is fixed at $-1$. From the change of the three Chern numbers, we can infer: 1) for $\theta>\theta_c$, the Chern numbers for the remote bands below bottom gap is 0 while for the flat bands are $\pm1$; 2) for $\theta<\theta_c$, the Chern number for the remote bands is $-1$ while for the flat bands are 0.  }

\begin{figure}
  \begin{center}
  \includegraphics[width=8.5 cm]{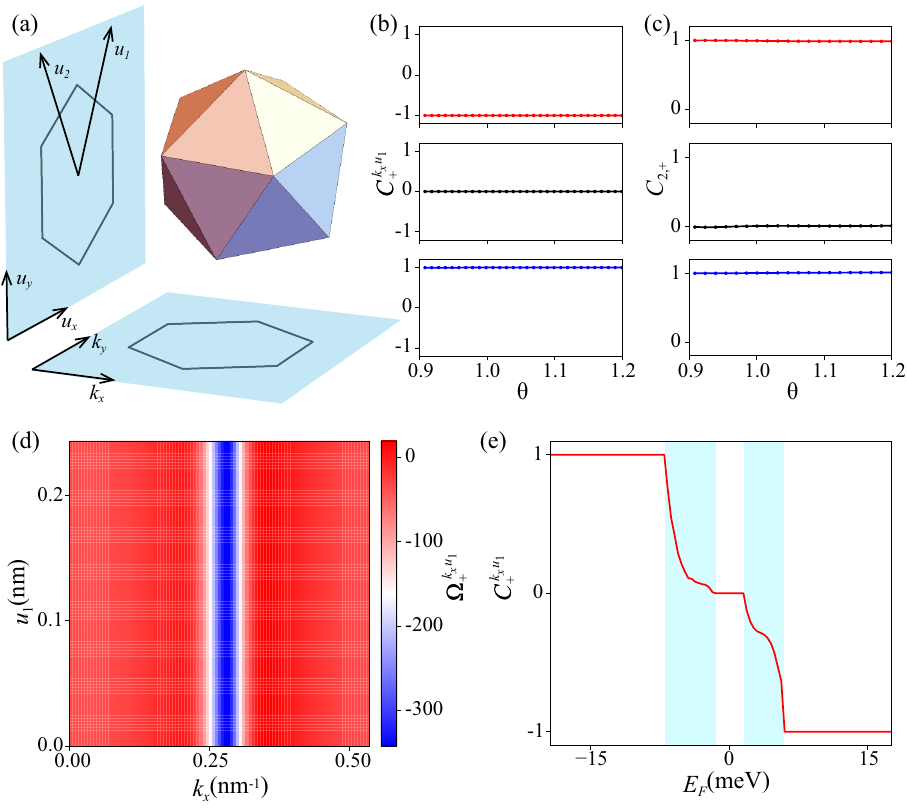}
  \end{center}
\caption{Schematic 4D BZ and its projection onto the $k_x$-$k_y$ and $u_x$-$u_y$ planes.  The DCN $C_+^{k_x u_1}$ (b) and the second Chern number $C_{2,+}$ (c) as a function of the twist angle $\theta$. The blue, black, red lines correspond the Fermi energy in the bottom, middle, top gaps in Fig. \ref{fig3}(b). (d) The Berry curvature $\Omega_+^{k_xu_1}$ for the lower flat band and with $k_y=0$. (e) The DCN $C_+^{k_x u_1}$ as a function of Fermi energy $E_F$. 
} 
  \label{fig4}
\end{figure}

{Now we consider the sliding of the top layer graphene by $\bm{u}$ as shown in Fig. \ref{fig1}(b).}
Because the Hamiltonian  Eq. (\ref{HTBG}) is a periodic function of both $\bm{k}$ and $\bm{u}$, the sliding TBG resides on a compact 4D manifold spanned by $(k_x,k_y,u_x,u_y)$.  The effective BZ in the 4D space and its projection onto the orthogonal $k_x$-$k_y$ plane and $u_x$-$u_y$ plane are schematically shown  in Fig. \ref{fig4}(a). {In the $u_x$-$u_y$ plane, the effective projected BZ is just the unit cell of top layer graphene. In the compact 4D manifold, the DCN can be defined on arbitrary 2D hyperplanes associated with synthetic dimensions. Without loss of generality, we slide the top layer graphene along the $u_1$ direction defined in Fig. \ref{fig4}(a) and the DCN $C_\tau^{k_xu_1}$ (for $\tau K$ valley) is defined in the $k_x$-$u_1$ plane. To study the interplay between twisting and sliding, 
$C_+^{k_xu_1}$ as a function of the twist angle $\theta$ is shown in Fig. \ref{fig4}(b). Here the blue, black, and red lines are respectively for the Fermi energy in the bottom, middle, and top band gaps in Fig. \ref{fig3}(b). The nontrivial topology manifested by the nonzero DCN $C_+^{k_xu_1}=\pm1$ for the Fermi energy in the bottom and top gap is independent of $\theta$ within the range. Moreover, we find DCN $C_+^{k_xu_1}=C_-^{k_xu_1}$. To understand this,}
 we  analyze the relation between Berry curvatures in two valleys. The Hamiltonian for the opposite valley is related by time reversal operation $\mathcal{H}_+(\bm{k}, \bm{u})=\mathcal{T} \mathcal{H}_-(-\bm{k}, \bm{u})\mathcal{T}^{-1}$ where $\mathcal{T}=\mathcal{K}$ is the complex conjugation operator. Then its eigenvalues and wavefunctions follow $E_{+,n}(\bm{k}, \bm{u})=E_{-,n}({-\bm{k}, \bm{u}})$ and $\phi_{+,n}(\bm{k}, \bm{u})=\phi_{-,n}^*(-\bm{k}, \bm{u})$. Therefore the Berry curvature obeys $\Omega_{+, n}^{k_xk_y}(\bm{k}, \bm{u})=-\Omega_{-,n}^{k_xk_y}(-\bm{k},\bm{u})$ according to the definition. This is the reason why the Chern number in the $k_x$-$k_y$ plane changes sign for the opposite valleys. Interestingly, the Berry curvatures associated with synthetic dimension in the plane spanned by $(k_\mu, u_\nu)$ follows $\Omega_{+, n}^{k_\mu u_\nu}(\bm{k},\bm{u})=\Omega_{-, n}^{k_\mu u_\nu}(-\bm{k},\bm{u})$ that yields $C_{+}^{k_\mu u_\nu}=C_{-}^{k_\mu u_\nu}$. {Moreover, due to incommensurate nature of the MH, the Berry curvature is dispersionless against $\bm{u}$ as shown in Fig. \ref{fig4}(d).} 

The DCN follows an addition rule that is, for the top layer graphene sliding along the lattice vector $\bm{u}_\nu =m \bm{u}_1 + n\bm{u}_2 $, the Chern number is $C_\tau^{k_\mu u_\nu}=mC_\tau^{k_\mu u_1}+nC_\tau^{k_\mu u_2}$. This is because $\ket{\partial_{u_\nu}\phi_{\tau,n}}=\ket{\nabla \phi_{\tau,n}} \cdot \bm{u}_\nu$ and hence $\Omega_{\tau,n}^{k_\mu u_\nu}=m \Omega_{\tau,n}^{k_\mu u_1} + n \Omega_{\tau,n}^{k_\mu u_2}$. 
According to the addition rule, the DCN for the  graphene layer sliding along arbitrary direction can be obtained from $C_\tau^{k_\mu u_1}$ and $C_\tau^{k_\mu u_2}$.

{The topological property of the 4D manifold is characterized by the second Chern number }
\begin{equation}
C_{2,\tau}=\frac{1}{4\pi^2}\sum_{E_n\le E_F}\int d k_x  d k_y  d u_1  d u_2 F_{\tau,n}^{k_xk_yu_1u_2},
\end{equation}
{where $F_{\tau,n}^{k_xk_yu_1u_2}= \Omega_{\tau,n}^{k_x k_y } \Omega_{\tau,n}^{u_1 u_2}+\Omega_{\tau,n}^{k_x u_1 } \Omega_{\tau,n}^{u_2 k_y}+\Omega_{\tau, n}^{k_x u_2 } \Omega_n^{k_y u_1 }$ \cite{yang1978generalization}. The second Chern number $C_{2,+}$ as a function of $\theta$ is displayed in Fig. \ref{fig4}(c). $C_{2,+}$ for the Fermi energy in the bottom to top gaps in Fig. \ref{fig3}(b) are respectively $1$, 0, and $1$ and independent of $\theta$. Moreover we find $C_{2,+}=C_{2,-}$ because $F_{\tau,n}^{k_xk_yu_1u_2}(\bm{k},\bm{u})=F_{\tau,n}^{k_xk_yu_1u_2}(-\bm{k},\bm{u})$ with $\Omega_{+, n}^{u_1u_2}(\bm{k}, \bm{u})=-\Omega_{-,n}^{u_1u_2}(-\bm{k},\bm{u})$. 
} {We remark that a nonzero DCN does not require to open a gap between the flat bands.}

The nontrivial topology generated dynamically can be measured through the electronic response of MHs to the sliding motion. The dynamics of electron due to the sliding (and without electromagnetic field) follows the semi-classical equation of motion 
\begin{equation}\label{de}
\dot{r}^{\mu}_{\tau,n}  = \frac{\partial E_{\tau, n}  (\bm{k},\bm{u}) }{\hbar \partial k_\mu}-\dot{u}_{\nu }  \Omega_{\tau,n}^{\mu \nu }(\bm{k},\bm{u}), 
\end{equation}
where $r^{\mu}_{\tau,n}$ is the center  of mass of the electronic wave packet \cite{RevModPhys.82.1959,4DQH}. The current induced by the sliding motion is 
\begin{equation}
j^\mu = e \sum_{E_n\le E_F,\tau} \int \frac{d^D \bm{k}} {(2\pi)^D} \dot{u}_{\nu }  \Omega_{\tau,n}^{\mu \nu }(\bm{k},\bm{u}),
\end{equation}
where  $D$ is the real dimension of the system. The contribution from the group velocity in Eq. (\ref{de}) vanishes because the dispersion $E_{\pm, n}$ of the two valleys are related by TRS. Thus the current is purely from the nontrivial {Berry curvature} generated by the sliding. The quantized topological charge pumping can be measured by integrating the current over a period of time of the sliding motion 
\begin{equation}
\begin{split}
q^\mu &= e \sum_{E_n\le E_F,\tau} \int \frac{d^D \bm{k} du_\nu} {(2\pi)^D}  \Omega_{\tau,n}^{\mu \nu } = e \sum_{\tau} \int \frac{d^{D-1} \bm{k}}{(2\pi)^{D-1}} C_{\tau}^{\mu\nu}.
\end{split}
\end{equation}
The contributions from the two valleys are identical for TBG.

We calculate explicitly the current {$I_l=-iet /\hbar \langle c_{i+1,l}^{\dagger} c_{i,l}-c_{i+1,l}^{\dagger} c_{i,l}\rangle$} through a bond in the bi-chain model by solving the time dependent Schr\"{o}dinger equation with the Fermi energy fixed in the band gap in Fig. \ref{fig2}(b). Here we shift the top chain adiabatically as $u=vt$ with the velocity $v=a_2t/50\hbar$. Current generated by the sliding in both chains are shown in Fig. \ref{fig2}(d). The current in the bottom chain is almost 0, while there is a net current in the top chain as expected. 
By integrating the current over one period, 
the pumped charge is $-1.08e$ that {agrees with the DCN}. When the MH is metallic, the Chern number is not quantized but the topological pumping persists. For example, the DCN $C_+^{k_x u_1}$ as function of the Fermi energy $E_F$ for the magic-angle TBG {is displayed} in Fig. \ref{fig4}(e). When $E_F$ cuts the flat bands in the shadowed regions, the DCN is fractional, which indicates that the number of pumped charge in one period of sliding is not quantized to integers. The nonlinear Hall response due to the nonzero $C_{2,\tau}$ can be possibly induced by shearing the MHs. \cite{4DQH}

To summarize, {we show that} nontrivial topology is generated dynamically by interlayer sliding in MHs. 
The nontrivial topology is characterized by the first and second Chern numbers in the high dimensional manifold associated with the synthetic dimensions induced by the sliding. The interplay between the Chern numbers and twist angle shows that even though the electronic state is topologically trivial in real dimensions, it can acquires a nonzero DCN in the synthetic dimensions. Due to the nontrivial topology, there is topological pumping caused by the sliding motion that can be used to measure the DCN. The sliding bi-chain model can possibly be realized in the double-wall carbon nanotube \cite{zhao2019observation}, while the dynamical control of twisting and sliding in 2D MHs has already been achieved in experiments \cite{Ribeiro-Palau690}. 
{The sliding MHs thus provide a new platform to study dynamically generated topology. \cite{Zhang_Four_2001,Kraus_Four_2013,Price_Four_2015,Lohse_Exploring_2018,Zilberberg_Photonic_2018,Su_Nontrivial_2018,Petrides_Six_2018}}

\begin{acknowledgements}

Computer resources for numerical calculations were supported by the Institutional Computing Program at LANL. This work was carried out under the auspices of the U.S. DOE NNSA under contract No. 89233218CNA000001 through the LDRD Program, and was supported by the Center for Nonlinear Studies at LANL. This work was also supported by the U.S. Department of Energy, Office of Science, Basic Energy Sciences, Materials Sciences and Engineering Division, Condensed Matter Theory Program.

\end{acknowledgements}

{\it Note added: ---} While finalizing the manuscript, we become aware of Ref. \cite{fujimoto2019topological}, which has some overlap with the present work.

\bibliography{references}

\end{document}